**easylayout: an R package for interactive force-directed layouts within RStudio**

Danilo Oliveira Imparato[1], João Vitor F. Cavalcante[1], Rodrigo J. S. Dalmolin[1]

[1]Bioinformatics Multidisciplinary Environment, Federal University of Rio Grande do Norte, Natal, Brazil

ABSTRACT

**Motivation**

Network visualization is critical for effective communication in various fields of knowledge. Currently, a gap separates network manipulation from network visualization in programming environments. Users often export network data to be laid out in external interactive software, like Cytoscape and Gephi. We argue the current R package ecosystem lacks an interactive layout engine well integrated with common data analysis workflows.

**Results**

We present easylayout, an R package that bridges network manipulation and visualization by leveraging interactive force simulations within the IDE itself (e.g., RStudio, VSCode). It is not yet another visualization library, but instead aims to interconnect existing libraries and streamline their usage into the R ecosystem. easylayout takes an igraph object and serializes it into a web application integrated with the IDE's interface through a Shiny server. The web application lays out the network by simulating attraction and repulsion forces. Simulation parameters can be adjusted in real-time. An editing mode allows moving and rotating nodes. The implementation aims for performance, so that even lower-end devices are able to work with relatively large networks. Once the user finishes tweaking the layout, it is sent back to the R session to be plotted through popular libraries like ggraph, igraph or even the base package itself. The current implementation focuses on the R ecosystem, but using web technologies makes it easily portable to similar environments, like Python/Jupyter Notebooks. We expect this tool to reduce the time spent searching for suitable network layouts, ultimately allowing researchers to generate more compelling figures.

**Availability and implementation**

easylayout is freely available under an MIT license on GitHub (https://github.com/dalmolingroup/easylayout). The package is implemented in R/Shiny and JavaScript/Svelte.

**INTRODUCTION**

Network studies have applications in various fields of knowledge. Visualizing networks is an important step in consolidating these studies, as it provides a quick and intuitive way to understand their results. The node-link diagram is the most common way to visualize a network. In this diagram, the system components (nodes or vertices) are depicted as points, and the interactions between them (edges) are represented by lines. Network diagrams use layout algorithms to position vertices and edges within a coordinate system, visually conveying the relationships between the represented entities (Kwon & Ma, 2020).

One of the most popular classes of graph layout algorithms is the force-directed, which simulates mechanical systems where vertices are particles that repel each other, and edges act as attractive forces between them (Kobourov, 2012). As the system evolves over time, edges reach a uniform length, and unconnected vertices move farther apart.

Different algorithms and simulation parameters produce layouts that emphasize different aspects of the same network. This means users often need to test multiple combinations of algorithms and parameters before arriving at a satisfactory layout (Kwon & Ma, 2020). An advantage of using force-directed algorithms is the ability to visualize the simulation in real-time and interactively adjust its parameters to achieve more desirable layouts (Jacomy et al., 2014).

Before generating a layout, the data underlying the networks are usually manipulated in programming environments, such as the R language. The R community maintains packages that structure a wide range of analyses across all fields of knowledge (R Core Team, 2024; Giorgi et al., 2022). The igraph package is widely used for graph analysis due to its computational efficiency and extensive functionality, including layout and visualization algorithms (Csárdi et al., 2024; Csardi & Nepusz, 2006). However, igraph's layout algorithms have weaknesses, such as lacking interactivity, not incorporating more recent algorithms, and having implementation limitations like failing to prevent vertex overlap. As a result, users often export their data out of the R environment to continue the visualization manually in external software, increasing the surface area for reproducibility issues.

When it comes to network visualization, software such as Cytoscape and Gephi serve multiple functions: they not only compute layouts but also map visual information to graph elements. For example, in protein-protein interaction networks, the expression level of a protein can be mapped to the color of its corresponding vertex (Shannon et al., 2003; Bastian et al., 2009). In fact, these softwares have so many built-in network analysis utilities they be considered fully featured graph theory toolboxes. Cytoscape even goes a step further by integrating with biological databases out-of-the-box. This manual approach contrasts with the R package ecosystem, long praised for its automation and graphical capabilities. Moreover, layout algorithms are not dependent on specific visualization software and can be executed in various ways.

**The Current Ecosystem of R Packages for Network Layouts**

Several R packages offer network layout capabilities, each with distinct features and limitations. In this section, we briefly discuss some of popular packages and how they relate to the present study.

We loosely categorize packages based on whether or not they provide real-time control of parameters while the algorithm iterates. Static tools, by design, merely output final node coordinates once, dismissing intermediate steps. That is, no user interaction happens during computation. In contrast, interactive tools can animate node positions as the algorithm iterates, allowing users to adjust simulation parameters in real-time and influence the resulting layout dynamically. They can run and animate for as long as the user wishes.

### graphlayouts

The graphlayouts package introduces layout algorithms previously unavailable in R, complementing the set of algorithms already available in igraph. It adds one fast general-purpose algorithm and others that emphasize group structures or the positioning of individual nodes (Schoch, 2023). This package is limited to static layouts.

### ggraph

The main focus of the ggraph package is extending ggplot2 to work with graph objects. Besides importing all layouts from graphlayouts, it introduces even more specialized options such as hive plots and biofabric layouts (Pedersen, 2017). It is also limited to static layouts.

**RCy3**

RCy3 communicates with a running instance of the Cytoscape desktop application via a REST API, effectively automating its usage from within R (Gustavsen et al., 2019). It provides R functions to all features available in the desktop application, which includes powerful layout features, like multiple algorithms, node alignment, even spacing, and component packing. Since RCy3 is merely an automation API for the desktop application, it inherits the former's limitations.

**RCyjs and cyjShiny**

Both RCyjs and cyjShiny wrap the Cytoscape.js JavaScript library for use within R. Cytoscape.js, like the desktop app, is also a fully featured graph theory tool and can be used to render interactive graphs in the browser (Shannon, 2024; Luna et al., 2023; Franz et al., 2023). cyjShiny focuses on integrating with Shiny apps. A significant drawback is that Cytoscape.js currently lacks a WebGL renderer, hindering performance for larger networks (Höffner et al., 2024). Although it is possible to use these wrappers to extract layout coordinates back into the R session, there is no straightforward way to adjust layout parameters in real-time. Moreover, the Cytoscape ecosystem brings a lot of overhead from being a fully featured graph theory library. A lot of effort is put into aesthetics and interactivity, and not much on fine-tuning simulations in real-time.

**visNetwork**

visNetwork wraps all features of the vis.js JavaScript library to create browser visualizations within the IDE's viewer pane (Almende B.V. and Contributors & Thieurmel, 2015). The package focuses on aesthetic customizations and interactivity. It does not operate on igraph objects, but rather introduces a new class. The documentation has instructions on how to use igraph layouts instead of vis.js own physics engine. In other words, information flows from igraph to visNetwork by design, but not easily the other way around, diverting the developer experience from the igraph package into its own. Users can export to PNG or HTML, which has limited use for academic publications.

**networkD3**

networkD3 leverages the D3.js JavaScript library to also display networks within the viewer pane (Allaire et al., 2014). It supports various layouts algorithms besides the traditional force simulation, like Sankey diagrams and dendrograms. It also does not operate on igraph objects, requiring conversion into a specific list structure. Exports are limited to HTML.

**Summary of Limitations**

These packages offer valuable features for network visualization in R, but they share common limitations, such as: lack of interactivity; low performance when interactive; lack of two-way interoperability with igraph; and dependency on external software.

While graphlayouts represents a step forward in shedding feature overhead, we still lack the flexibility of tweaking force simulations in real-time. An interactive layout engine within the programming environment is a missing piece in the R network analysis workflow.

**The Current Ecosystem of R Packages for Plotting Networks**

    **igraph**

The igraph package provides the plot.igraph function, which non-interactively plots 2D node-link diagrams of igraph objects. It is built on top of base R packages and therefore can output to a multitude of formats, like PDF, SVG, and PNG. It provides detailed control over the appearance of network elements, including node attributes (e.g., width, height, shape, fill color, border properties) and edge attributes (e.g., color, thickness, arrow types), to name a few (Csárdi et al., 2006). The plotting flexibility is remarkable, and should suffice most use cases.

    **ggraph**

ggraph extends ggplot2 to support network by introducing custom geoms and, importantly, facets (Pedersen, 2017). As discussed previously in this paper, it also introduces new layout algorithms. It is by far the most popular and feature-rich ggplot2 extension for networks, although other packages do exist (Briatte, 2016). A key advantage is allowing users to plot networks within the tidy paradigm (Wickham et al., 2019; Pedersen, 2017b). Furthermore, it brings even more advanced graphical capabilities. For instance, ggraph implements different types of edge bundling, edge color gradients, edge loops, tile nodes, and voronoi nodes, to name a few. Notably, ggraph synergizes with the ggforce package, from the same author, which introduces features like label overlap prevention, and partial zooming.

### ggplot2 and base graphics

ggraph and igraph generally meet the requirements of most network visualization tasks. However, there are scenarios where more granular control is necessary. In such cases, users can resort to ggplot2 or base R graphics, which are the primary tools for general-purpose plotting in R (Wickham, 2016). Under the philosophy of igraph, ggraph, graphlayouts, and easylayout, a layout is simply a matrix of X and Y coordinates. A minimal plot can be assembled by plotting points and lines based on those coordinates. More advanced aesthetics can be progressively added on top of that, according to users' needs. In fact, our group often chooses to plot networks directly with ggplot2, due to some advanced facetting requirements (Viscardi et al., 2020). As far as we are aware, this amount of control cannot be as easily leveraged with any other available tool, even outside R.

## R Packages That Served as Inspiration

Several R packages inspired the development of this present work, all of which share a common theme: they augment developer experience by introducing a UI within the IDE itself. The information flows back to R from the UI just as easily as it came in. This ensures the UI does not act as a dead ends in the user's workflow. They keep R code as the authoritative medium for analysis, offering a consistent developer experience and promoting reproducibility. Notably, these packages are all closely tied to ggplot2.

### esquisse

esquisse is an RStudio addin designed for interactive data exploration and visualization using ggplot2 (Meyer & Perrier, 2024). It allows users to create diverse visualizations, such as bar plots, scatter plots, histograms, and boxplots, while also exporting the code used to create the plots. The code export tool is arguably the most valuable feature of the package. It enables even inexperienced users to generate R code for high-quality graphics through drag-and-drop interactions. This approach effectively teaches users the underlying code while they create visualizations. In other words, the end result is not the plot itself, but rather the code to generate it, promoting reproducibility.

### snahelper

snahelper offers a set of RStudio addins for network analysis (Schoch, 2024). The main addin, provides GUI for performing common network analysis tasks and visualizing networks with ggraph. It also includes addins for: visually building networks and exporting them as igraph objects; 2) importing network data and generating the code for it; 3) manually packing individual network components and persisting the adjustments to the original igraph object.

### ggiraph

ggiraph creates interactive ggplot2 graphics. It adds features such as tooltips, hover effects, selection, and custom JavaScript actions (Gohel & Skintzos, 2024). It works by exporting plots as SVG elements with custom attributes, allowing JavaScript to add interactivity to an SVG DOM. Importantly, when used within a Shiny application, interacted elements become available to the R backend.

**Our proposal**

We present easylayout, an R package that interactively generates network layouts using force simulations within the IDE itself (e.g., RStudio, VSCode). It is not yet another visualization library, but rather an effort to standardize and interconnect existing libraries. The package bridges network analysis and visualization within the same environment, improving convenience, control, and reproducibility.

**IMPLEMENTATION**

**Choice of Libraries for Force Simulation**

Interactive force simulation is the main focus of the package and was implemented using JavaScript libraries for the web. Most popular IDEs are built with web technologies, such as RStudio, VSCode, and Jupyter Notebook. Choosing a web library becomes an obvious choice given the ease of integration with these tools.

**VivaGraphJS**

VivaGraphJS is a JavaScript library used to simulate forces and visualize graphs in the browser (Kashcha, 2024). Its features are divided into different modules. The modular structure allows developers to replace or extend components using different libraries or custom code. Another key feature is its computational efficiency, which enables smoother simulations compared to similar libraries. Based on its modularity and performance, we chose VivaGraphJS as the starting point for developing easylayout.

**d3-force**

VivaGraphJS implements a single force simulation algorithm, whereas tools like Gephi and Cytoscape offer multiple options. Different algorithms come with different parameters which suit different visualization requirements. We decided to include d3-force, the most popular force algorithm library for the web. D3-force is a module of the D3.js library, the gold standard for advanced data visualizations in the browser (Bostock et al., 2011). Although more computationally intensive than VivaGraphJS, its popularity and wide range of adjustable parameters were strong reasons to integrate it into easylayout [10].

**Integration with the IDE**

easylayout leverages the R package ecosystem to run force simulations directly within the RStudio interface. The "Viewer pane" in RStudio allows web pages to be opened directly within the IDE. For example, instead of generating a static image of a scatterplot, users can opt to generate a web page that adds interactivity to the plot.

One way to take advantage of the Viewer pane's convenience is through the Shiny package. Shiny enables the development of full web applications by running a server with an R-based backend (Chang et al., 2024). Notably, it provides tools for transmitting data between the R environment and the web application. Easylayout relies on this feature to transmit an igraph object into a web application responsible for running the force simulation. The web app is automatically loaded in the Viewer pane once our package is run.

The final output of the simulation is an XY coordinate matrix with N rows, where N represents the number of vertices in the graph. This layout format is compatible with igraph and other related packages.

**Implementation of the Interactive Simulation Module**

The interactive simulation module corresponds to the web app that handles: 1) loading a JSON sent by the Shiny server, 2) converting it into an object compatible with the VivaGraphJS library, 3) rendering a simulation on the screen, and 4) providing a specialized interface to control the simulation parameters in real-time.

The app was developed with the Svelte 4 JavaScript framework. Svelte is known for its simplicity and ease of use when compared to alternative frameworks. The simple syntax lowers the entry barrier for developers with less experience in web technologies (Svelte: Cybernetically Enhanced Web Apps, 2024; Bhardwaz & Godha, 2023). Moreover, it is also the most performant choice among popular modern web frameworks (Ollila et al., 2022).

VivaGraphJS has its own implementation of a graphics engine based on WebGL. WebGL is an API available in browsers that allows images to be rendered using the GPU, making it the most performant way to render the simulation.

**Implementation of the Editor Module**

Besides force simulation, it is also useful to allow users to manually move vertices by dragging them with the mouse. For instance, the repulsion exerted by the network's core often pushes peripheral vertices too far apart. This excessive distancing of certain parts of the network alters the visualization scale and compromises the final figure. It's common that no combination of parameters can aesthetically lay out the all parts of the network at once. In other cases, the distance between the vertices might be pleasing, but the layout as a whole adopts an oblique orientation, making it harder to arrange a final figure for publication. In such situations, arbitrarily rotating the layout becomes necessary.

Drag and rotation features could have been implemented on top of the simulation WebGL rendering context. However, implementing interactions for WebGL renderings is notoriously difficult. Therefore, drag and rotation were implemented using the Fabric.js library (fabric.js, 2024). Fabric.js offers a wide range of features for geometric manipulation of objects but lacks support for WebGL rendering. This means that rendering the editor mode is computationally more intensive.

**RESULTS**

The implementation of easylayout demonstrates it is possible to combine the steps of network analysis and visualization without relying on software external to RStudio.

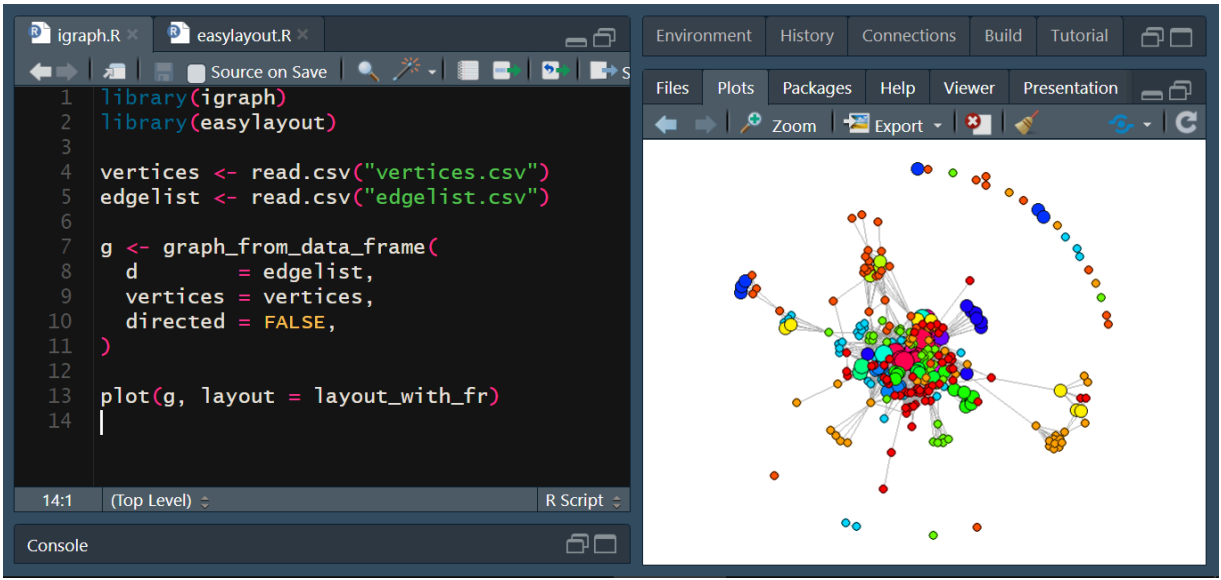

**Figure 1:** Plotting a network with igraph only. The layout algorithm is Fruchterman-Reingold. Nodes overlap and loose components drift away.

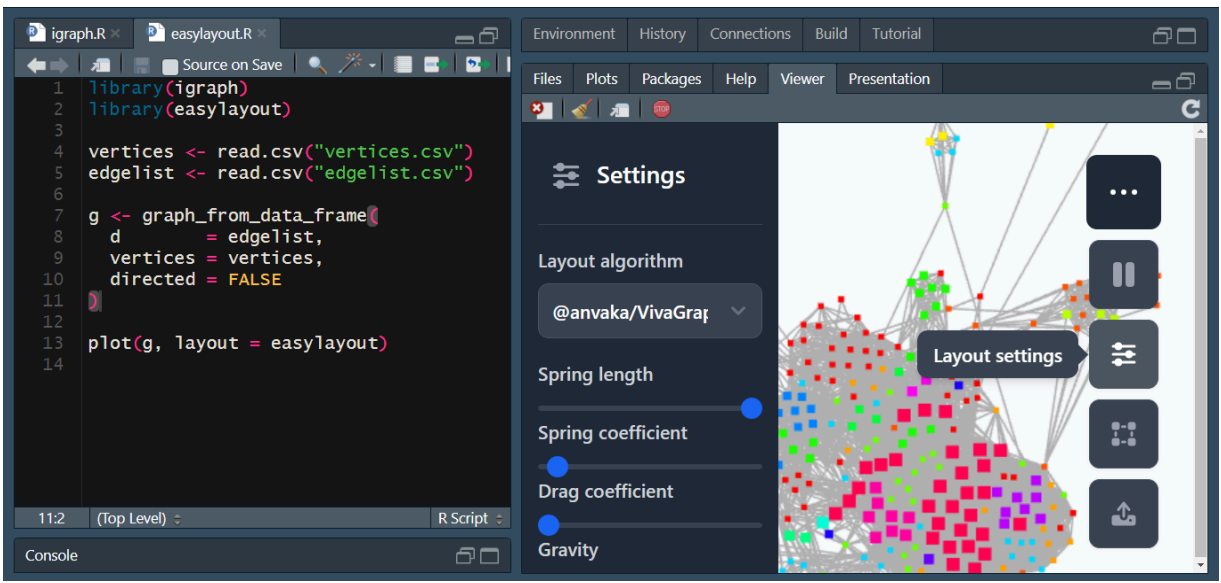

**Figure 2:** Plotting a network with igraph and easylayout. The "layout = easylayout" in line 13 is the only difference from the previous code. R code execution hangs and the Viewer pane opens automatically, exposing the web app which renders the network for simulation. The UI allows pausing and navigating between simulation and editor mode. A sidebar allows

adjusting force parameters in real-time. The "Finish" button on the bottom left corner terminates the app and resumes R code execution.

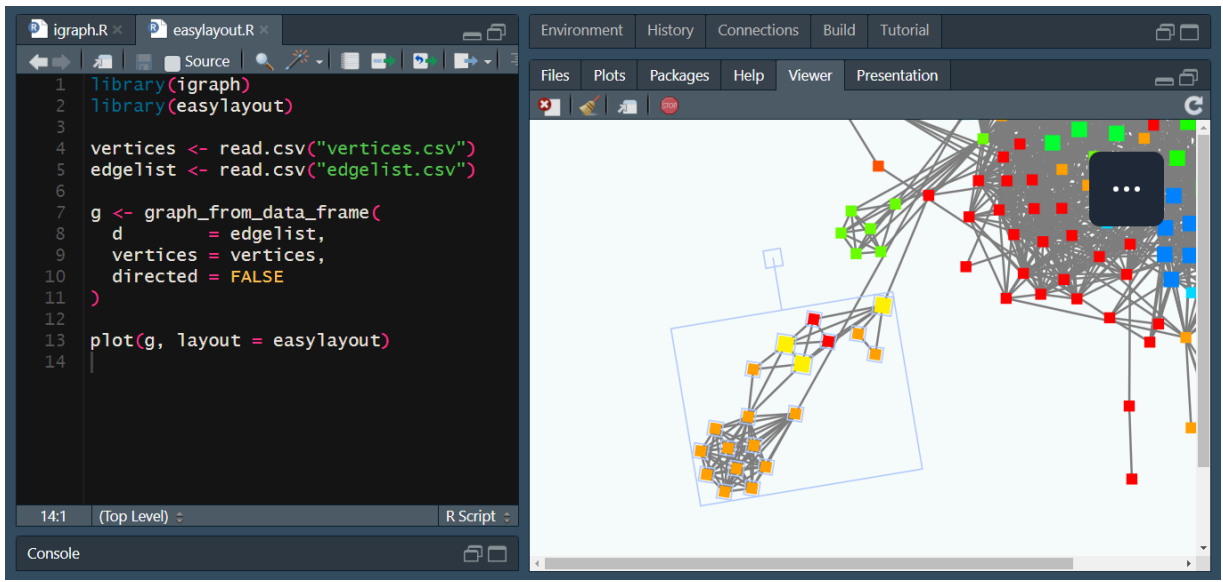

**Figure 3:** Editor mode. The user can select and drag around nodes and groups of nodes. Notably, the app also allows rotating groups of nodes together, as shown in the figure.

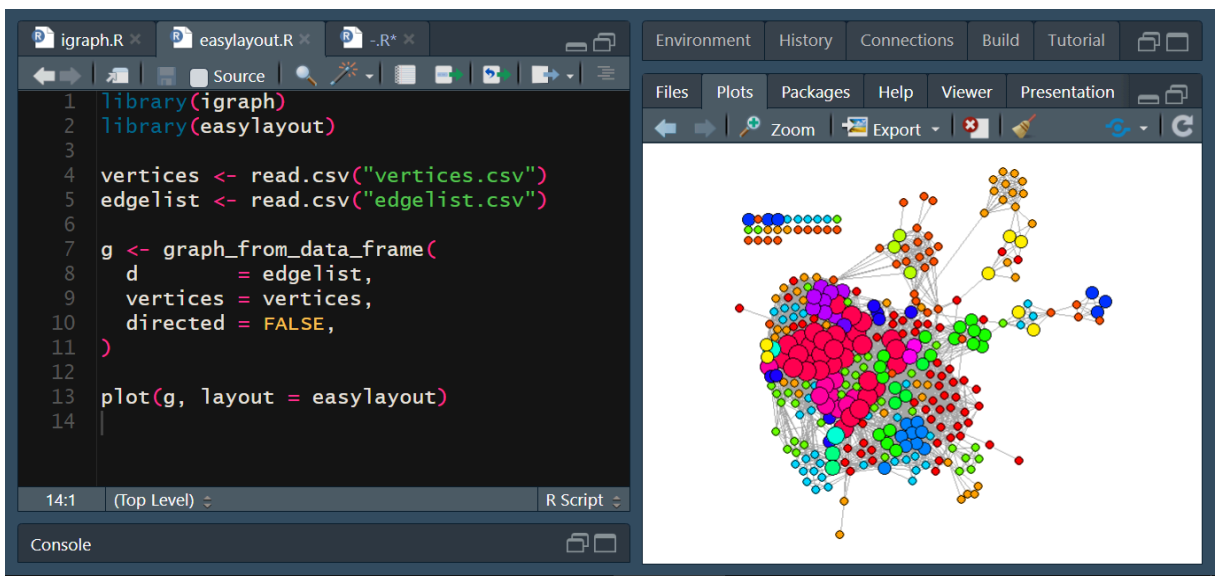

**Figure 4:** Since easylayout was called within the plot function itself, the network is plotted immediately after resuming code exection.

**CONCLUSION**

We hope easylayout improves the workflow for users and researchers working with network analysis and visualization. To achieve this goal, the package focuses on several key aspects: integration with the IDE, eliminating the need for external software; performance, by leveraging efficient libraries; harmony with the existing R package ecosystem; and ease of manually adjusting layouts.

On and ending note, we acknowledge that many users choose Python and Jupyter for their analyses, instead of R and RStudio. Since much of the easylayout package is implemented with web technologies, developing a Python version for Jupyter Notebook would require minimal effort.